\documentclass[useAMS,usenatbib]{mn2e}
\usepackage{epsfig}
\usepackage{hyperref}

\title[The first $\gamma$-ray detection of the narrow-line Seyfert 1 FBQS J1644$+$2619]{The first $\gamma$-ray detection of the narrow-line Seyfert 1 FBQS J1644$+$2619}
\author[F. D'Ammando, M. Orienti, J. Larsson, M. Giroletti]{F. D'Ammando$^{1,2}$\thanks{E-mail: dammando@ira.inaf.it},
  M. Orienti$^{2}$, J. Larsson$^{3}$, M. Giroletti$^{2}$\\
$^{1}$Dip. di Fisica e Astronomia , Universit\`a di Bologna, Via Ranzani 1, I-40127 Bologna, Italy \\
$^{2}$INAF-Istituto di Radioastronomia, Via Gobetti 101, I-40129 Bologna, Italy\\
$^{3}$KTH, Department of Physics, and the Oskar Klein Centre, AlbaNova, SE-106 91 Stockholm, Sweden \\
$^{4}$Hiroshima Astrophysical Science Center, Hiroshima University, 1-3-1 Kagamiyama, Higashi-Hiroshima 739-8526, Japan}

\begin{document}

\date{Accepted. Received; in original form}

\maketitle

\label{firstpage}

\begin{abstract}
We report the discovery of $\gamma$-ray emission from the narrow-line Seyfert 1 (NLSy1)
galaxy FBQS J1644$+$2619 by the Large Area Telescope on board the {\em Fermi}
satellite. The Third Fermi LAT Source
catalogue reports an unidentified $\gamma$-ray source, detected over the first
four years of {\em Fermi} operation, 0\fdg23 from the radio position of the NLSy1. Analysing 76 months of
$\gamma$-ray data (2008 August 4--2014 December 31) we are able to better constrain the localization of
the $\gamma$-ray source. The new position of the $\gamma$-ray source is 0\fdg05 from
FBQS J1644$+$2619, suggesting a spatial association with the NLSy1. This is
the sixth NLSy1 detected at high significance by {\em Fermi}-LAT so far. Notably, a significant increase of activity was observed in $\gamma$-rays
from FBQS J1644$+$2619 during 2012 July--October, and an increase of activity in
$V$-band was detected by the Catalina Real-Time Sky Survey in the same period. 
\end{abstract}


\section{Introduction}

The discovery of variable $\gamma$-ray emission from narrow-line Seyfert 1 (NLSy1) galaxies revealed the presence of a possible
third class of Active Galactic Nuclei (AGN) with relativistic jets \citep{abdo09}. Five NLSy1 have been detected at high significance in
$\gamma$-rays by the Large Area Telescope (LAT) on board the {\em Fermi Gamma-ray Space Telescope} satellite so far \citep{abdo09, dammando12}. 
High brightness temperatures and radio-loudness parameters have been observed in the NLSy1 already detected by {\em Fermi}-LAT. Therefore, these properties
may be a good proxy for the jet production efficiency, and thus for selecting
the best candidates for a LAT $\gamma$-ray detection. However, in the
radio-loud NLSy1 sample presented by \citet{yuan08}, neither the source with
the highest radio-loudness (B3 1044$+$476) nor the highest brightness
temperature (TXS 1546$+$353) has yet been detected in $\gamma$-rays. 

The NLSy1 detected in $\gamma$-rays so far have exhibited blazar-like behaviour, including strong $\gamma$-ray flux variability
\citep{dammando13,dammando14,foschini15}. Due to this variability and their
apparent rarity, identification of new $\gamma$-ray emitting NLSy1 will
benefit enormously from frequent observations of a large number of
candidates. The {\em Fermi}-LAT's nearly continuous surveying of the entire
$\gamma$-ray sky provides a unique opportunity to carry out such monitoring, with great
sensitivity. As the {\em Fermi} satellite continues to accumulate data, the
discovery of new NLSy1 in $\gamma$-rays is expected. 

The Third {\em Fermi} LAT source catalogue \citep[3FGL;][]{acero15} reports an
unassociated $\gamma$-ray source, 3FGL J1644.4$+$2632 (R.A.=251\fdg123,
Dec.=26\fdg542, with a 95 per cent error circle radius of 0\fdg201). The radio position of First Bright Quasar Survey (FBQS)
  J1644$+$2619 (R.A.=251\fdg177, Dec.=26\fdg320) lies 0\fdg227 from the
  $\gamma$-ray source, just outside its 95 per cent error circle radius. For this reason, 3FGL J1644.4$+$2632 was not associated with FBQS J1644$+$2619 in the 3FGL catalogue.
FBQS J1644$+$2619 is a NLSy1
at redshift $z$ = 0.145 \citep{bade95}, with a radio loudness\footnote{The radio-loudness $R$ is defined as the ratio of radio at 1.4 GHz to $B$-band flux densities.} of $R$ = 447, a full width at half-maximum
(FWHM) (H$\beta$) = (1507 $\pm$ 42)\,km\,s$^{-1}$, and [O III]/H$\beta$ $\sim$
0.1 \citep{yuan08}. In this paper we present results of the analysis of
76 months of {\em Fermi}-LAT data reporting the first detection in $\gamma$-rays of the
NLSy1 FBQS J1644$+$2619.
 
This paper is organized as follows. In Section 2, we report the {\em Fermi}-LAT data analysis and results. The data collected by the Catalina Real-time Transient Survey (CRTS\footnote{{\url{http://crts.caltech.edu/}}}) in $V$-band, together with the X-ray, UV, and optical data collected by {\em Swift} are presented in Section 3. In Section 4, we discuss the properties of the source and draw our conclusions.

Throughout the paper the quoted uncertainties are given at the 1$\sigma$ level, unless otherwise stated, and the photon indices are
parameterized as $dN/dE \propto E^{-\Gamma_{\nu}}$, where $\Gamma_{\nu}$ is
the photon index in the different energy bands. We adopt a $\Lambda$ cold dark
matter cosmology with $H_0$ = 71 km s$^{-1}$ Mpc$^{-1}$,
$\Omega_{\Lambda} = 0.73$, and $\Omega_{\rm m} = 0.27$ \citep{komatsu11}. The redshift of this
source,  $z =0.145$, corresponds to a luminosity distance d$_{\rm\,L} =  690$\
Mpc, where 1 arcsec corresponds to a projected size of  2.56 kpc.

\section{{\em Fermi}-LAT Data: Selection and Analysis}
\label{FermiData}

\begin{figure}
\centering
\includegraphics[width=7.5cm]{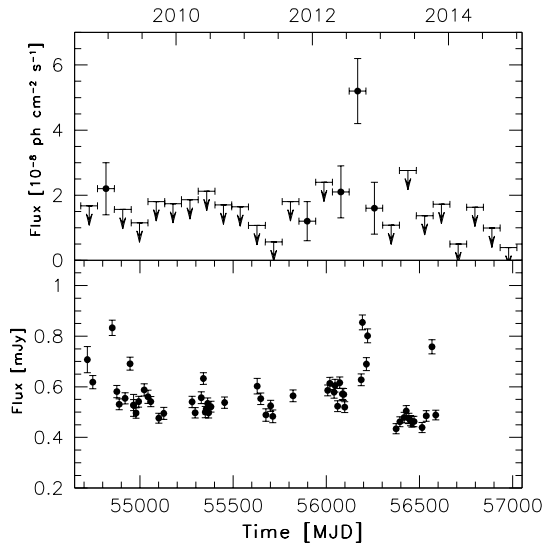}
\caption{{\it Upper panel}: integrated flux light curve of FBQS J1644$+$2619
  obtained by {\em Fermi}-LAT in the 0.1--100 GeV energy range during 2008
  August 4 -- 2014 December 31 (MJD 54682--57022) with 90-day time
  bins. Arrows refer to 2$\sigma$ upper limits on the source flux. Upper
  limits are computed when TS $<$ 5. {\it Lower panel}: light curve in $V$-band obtained by CRTS during 2008 October 8 -- 2013 October 22.}
\label{Fig1}
\end{figure}

The {\em Fermi}-LAT  is a pair-conversion telescope operating from 20 MeV to
$>$ 300 GeV. Details about the {\em Fermi}-LAT are given in \citet{atwood09}. 
The {\em Fermi}-LAT data reported in this paper were collected from 2008 August 4 (MJD
54682) to 2014 December 31 (MJD 57022). During this time the LAT instrument
operated most of the time in survey mode, scanning the entire sky every 3 hours. The analysis was performed with the
\texttt{ScienceTools} software package version v9r34p03. Only events belonging
to the `Source' class were used. The time intervals when the rocking angle of the LAT was greater than 52$^{\circ}$
were rejected. In addition, a cut on the zenith angle ($< 100^{\circ}$) was applied to reduce contamination from
the Earth limb $\gamma$-rays, which are produced by cosmic rays interacting with the upper atmosphere. 
The spectral analysis was performed with the instrument response functions \texttt{P7REP\_SOURCE\_V15} using an unbinned maximum likelihood method implemented in the Science tool \texttt{gtlike}. Isotropic (`iso\_source\_v05.txt') and Galactic diffuse
emission (`gll\_iem\_v05\_rev1.fit') components were used to model the background\footnote{http://fermi.gsfc.nasa.gov/ssc/data/access/lat/\\BackgroundModels.html}. The normalization of both components was allowed to vary freely during the spectral fitting.

We analysed a region of interest of $10^{\circ}$ radius centred at the
location of FBQS J1644$+$2619. We evaluated the significance of the $\gamma$-ray signal from the source by
means of a maximum-likelihood test statistic (TS) that results in TS = 2 (log$L_1$ - log$L_0$), where
$L$ is the likelihood of the data given the model with ($L_1$) or without
($L_0$) a point source at the position of FBQS J1644$+$2619
\citep[e.g.,][]{mattox96}. The source model used in
\texttt{gtlike} includes all the point sources from the 3FGL catalogue that
fall within $15^{\circ}$ of FBQS J1644$+$2619. The spectra of these sources
were parametrized by a power-law (PL), a log-parabola (LP), or a super
exponential cut-off, as in the 3FGL catalogue. 

We tested whether two distinct $\gamma$-ray sources (one at the radio position of FBQS J1644$+$2619 and one
at the $\gamma$-ray position of 3FGL J1644.4$+$2632) are detected
simultaneously by {\em Fermi}-LAT over 76 months of observations, including
both sources in the model. In the fitting procedure, the normalization
factors and the spectral parameters of the sources lying within 10$^{\circ}$ of FBQS J1644$+$2619 were left as free
parameters. For the sources located between 10$^{\circ}$ and 15$^{\circ}$ from
our target, we kept the normalization and the spectral parameters fixed to the values
from the 3FGL catalogue.
A first maximum likelihood analysis was performed yielding TS = 2 for 3FGL
J1644.4$+$2632 and TS = 20 for FBQS J1644$+$2619. We concluded that only one
$\gamma$-ray source at a position compatible with FBQS J1644+2619 is detected
by {\em Fermi}-LAT over 76 months.

We removed 3FGL J1644.4$+$2632 and all sources with TS $<$ 5. 
A second maximum likelihood analysis was performed
on the updated source model. Integrating over the period 2008 August 4 -- 2014
December 31 using a PL model, $dN/dE \propto$ $(E/E_{0})^{-\Gamma}$, a TS of
26 ($\sim$4.4$\sigma$) was obtained for FBQS J1644$+$2619 in the 0.1--100 GeV energy
range, with an average flux of (5.9 $\pm$ 1.9)$\times$10$^{-9}$ ph
cm$^{-2}$ s$^{-1}$ and a photon index $\Gamma$ = 2.5 $\pm$ 0.2. This flux corresponds to an apparent
isotropic (rest-frame) $\gamma$-ray luminosity of 1.6$\times$10$^{44}$ erg s$^{-1}$. A $\gamma$-ray point source localization was performed using the \texttt{gtfindsrc} tool over the photons extracted during the entire period. The fit results in R.A. =
251.168$^{\circ}$, Dec. = 26.372$^{\circ}$ with a 95\,per cent error circle
radius of 0\fdg091. The $\gamma$-ray source lies at an angular separation of 0\fdg053 from the radio
position of FBQS J1644$+$2619, suggesting a spatial association
between the $\gamma$-ray source and the NLSy1 FBQS J1644$+$2619.

Figure~\ref{Fig1} shows the $\gamma$-ray light curve for the first 76 months of
{\em Fermi}-LAT observations of FBQS J1644$+$2619 using a PL model and 90-day
time bins. For each time bin, the spectral parameters of FBQS J1644$+$2619 and
of all sources within 10$^{\circ}$ of it were frozen to the values resulting from
the likelihood analysis over the entire period. If TS $<$ 5, the 2$\sigma$
upper limits were calculated. The systematic
uncertainty in the flux is dominated by the systematic uncertainty in the
effective area \citep{ackermann12}, which amounts to 10 per cent below 100
MeV, decreasing linearly with the logarithm of energy to 5 per cent between 316 MeV and 10 GeV, and increasing linearly with the logarithm
of energy up to 15 per cent at 1 TeV\footnote{http://fermi.gsfc.nasa.gov/ssc/data/analysis/LAT\_caveats.html}.

FBQS J1644$+$2619 was detected sporadically by the {\em Fermi}-LAT, with an increase of
activity during 2012 July 15--October 12. Running a $\gamma$-ray point source localization over
the photons extracted during the period 2012 June--2013 January, in which the
source was detected continuously, the fit results in R.A. = 251.156$^{\circ}$, Dec. = 26.373$^{\circ}$ with a 95\,per cent error circle radius of 0\fdg096. The $\gamma$-ray source lies at an angular separation of 0\fdg056 from the radio
position of FBQS J1644$+$2619. The position estimated in the high-activity period lies at an angular separation of 0\fdg011 from the position
 estimated over 76 months of LAT observations.

\begin{figure}
\centering
\includegraphics[width=7.5cm]{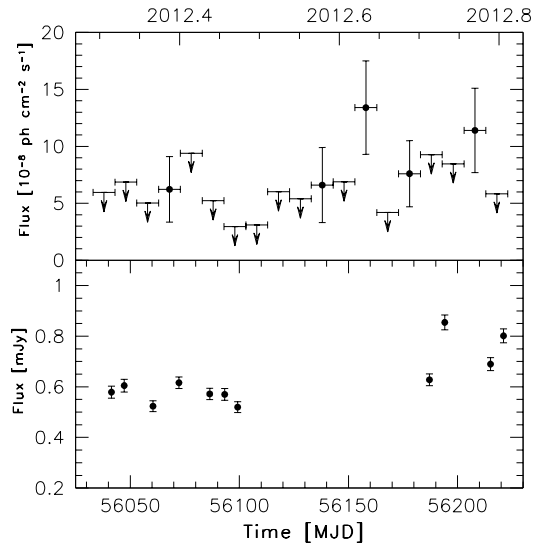}
\caption{{\it Upper panel}: integrated flux light curve of FBQS J1644$+$2619 obtained by {\em Fermi}-LAT in the 0.1--100 GeV energy range during 2012 April 16 -- October 22 (MJD 56033--56222) with 10-day time bins. Arrows refer to 2$\sigma$ upper limits on the source flux. Upper limits are computed when TS $<$ 5. {\it Lower panel}: light curve in $V$ band obtained by CRTS in the same period of the {\em Fermi}-LAT light curve.}
\label{Fig2}
\end{figure}

In the period 2012 July 15--October 12, the source reached a flux (0.1--100 GeV) of
(5.2 $\pm$ 1.0)$\times$10$^{-8}$ ph cm$^{-2}$ s$^{-1}$, a factor of 9 higher
than its average $\gamma$-ray flux. 
Leaving the photon index of our target
(and of all sources within 10$^{\circ}$ of our target) free to vary, the fit for FBQS
J1644$+$2619 results in a photon index $\Gamma_{\gamma}$ = 2.5 $\pm$ 0.2. Although limited by the statistics,
no significant spectral change seems to be detected during
the high state. A $\gamma$-ray light curve in the period 2012 April 16--October 22 with 10-day time bins is reported in Fig.~\ref{Fig2}.
On daily time-scales peak activity was detected on 2012 August 18 (MJD 56157) and 2012 October 5
(MJD 56205), with fluxes of
(66 $\pm$ 22)$\times$10$^{-8}$ ph cm$^{-2}$ s$^{-1}$ and
(43 $\pm$ 17)$\times$10$^{-8}$ ph cm$^{-2}$ s$^{-1}$, respectively. By means of the \texttt{gtsrcprob} tool, we estimated that the highest energy
photon detected from FBQS J1644$+$2619 was observed on 2012 June 6 at a
distance of 0\fdg002 from the $\gamma$-ray source with an energy of 25.7 GeV.

\section{Multifrequency data}

\subsection{Catalina Real-time Transient Survey data}

The source has been monitored by the CRTS \citep{drake09, djorgovski11},
using the 0.68-m Schmidt telescope at Catalina Station, Arizona, and an unfiltered
CCD. The typical cadence is 4 exposures separated by 10 min in a
given night; this may be repeated up to 4 times per lunation, over a period of
$\sim$6--7 months each year, while the field is observable. Photometry is
obtained using the standard Source-Extractor package \citep{bertin96}, and
transformed from the unfiltered instrumental magnitude to Cousins
$V$\footnote{\url{http://nesssi.cacr.caltech.edu/DataRelease/FAQ2.html\#improve}}
by $V$ = $V_{\rm CSS}$ + 0.31($B-V$)$^{2}$ + 0.04. We averaged the values obtained during the same observing night. 
During the CRTS monitoring the source showed a variability amplitude of $\sim$1 mag, changing between 17.78 and 16.72
mag. The CRTS flux densities, corrected for extinction using the E(B--V) value
of 0.073 from \citet{schlafly11} and the extinction laws from
\citet{cardelli89}, are reported in Figs.~\ref{Fig1} and \ref{Fig2}.

\subsection{{\em Swift} Data: Analysis and Results}
\label{SwiftData}

The {\em Swift} satellite \citep{gehrels04} observed FBQS
J1644$+$2619 on 2011 December 26 with the X-ray Telescope
\citep[XRT;][0.2--10.0 keV]{burrows05} and the Ultraviolet/Optical Telescope
\citep[UVOT;][170--600 nm]{roming05}. The XRT data were processed by using the \texttt{xrtpipeline
  v0.13.1} of the \texttt{HEASoft} package (v6.16) with standard procedures,
filtering, and screening criteria. The data were collected in photon counting
mode for a net exposure of about 1.3 ksec. Considering the low number of
photons collected ($<$ 200 counts) the spectra were rebinned with a minimum of 1 count per bin and we performed the fit with the Cash statistic \citep{cash79}. Source events were extracted from a circular region with a radius of
20 pixels (1 pixel $\sim$ 2.36 arcsec), while background events were extracted
from a circular region with radius of 50 pixels far away from bright sources. Ancillary response files were generated with \texttt{xrtmkarf}, and
account for different extraction regions, vignetting and point spread function
corrections. We used the spectral redistribution matrices in the
Calibration database maintained by HEASARC. The X-ray spectrum can be fitted
in the 0.3--10 keV energy range with an absorbed power-law using the photoelectric absorption model
\texttt{tbabs} \citep{wilms00}, with a neutral hydrogen column fixed to its Galactic value \citep[5.14$\times$10$^{20}$
cm$^{-2}$;][]{kalberla05} and a photon index of $\Gamma_{\rm\,X}$ =
2.0 $\pm$ 0.3 ($\chi^2$/d.o.f = 80/59). The
corresponding unabsorbed 0.3--10 keV flux is (2.3$\pm$0.5)$\times$10$^{-12}$
erg cm$^{-2}$ s$^{-1}$. Adding a thermal component (i.e. a black body model), as in
\citet{yuan08}, yields a temperature kT = 0.06$^{+0.05}_{-0.04}$ keV and a photon index $\Gamma_{\rm\,X}$ =
1.7$^{+0.4}_{-0.5}$ with a comparable quality of fit ($\chi^2$/d.o.f = 78/57). The low number of counts does not allow a detailed
comparison between the two models.

UVOT data in the $v$, $b$, $u$, $w1$, $m2$, and $w2$ filters were analysed
with the \texttt{uvotsource} task included in the
\texttt{HEASoft} package (v6.16) and the 20130118 CALDB-UVOTA release. Source counts were extracted from a circular region
of 5 arcsec radius centred on the source, while background
counts were derived from a circular region with 10 arcsec radius in a
nearby source-free region. The observed magnitudes are: $v$ = 17.71$\pm$0.24,
$b$ = 18.09$\pm$0.14, $u$ = 16.81$\pm$0.08, $w1$ = 16.96$\pm$0.08, $m2$ =
16.91$\pm$0.09, $w2$ = 16.88$\pm$0.05. Compared to the CRTS observations, UVOT detected the source in $V$-band during a low activity state.

\section{Discussion and Conclusions}

The confirmation of the existence of relativistic jets in NLSy1 opened a new
research path for improving our knowledge of the AGN phenomenon, but brought
new challenging questions. After the five objects detected by {\em Fermi}-LAT \citep{abdo09,dammando12}, no new
NLSy1 were reported in the 3FGL catalogue that covers the first 48 months of {\em Fermi} operation (2008 August 4--2012 July 31). The NLSy1 FBQS J1644$+$2619 lies 0\fdg23 from the $\gamma$-ray
source 3FGL J1644.4$+$2632, just outside its 95 per cent error circle of
0\fdg20. 3FGL J1644.4$+$2632 is a new $\gamma$-ray source detected at
4.5$\sigma$, not reported in the
previous {\em Fermi}-LAT catalogues \citep{abdo10,nolan12}, with a photon index of
$\Gamma_{\gamma}$ = 2.8 $\pm$ 0.2 and without an associated low-energy counterpart in the 3FGL catalogue. Analyzing 76 months of {\em Fermi}-LAT observations we were able to
better constrain the position of the $\gamma$-ray source and thus to have a
spatial association between the $\gamma$-ray source and the NLSy1 FBQS
J1644$+$2619. 

In Fig.~\ref{Fig1} we compare the {\em Fermi}-LAT light curve over the
period 2008 August--2014 December with the light curve obtained by CRTS in
$V$-band for FBQS J1644$+$2619. It is worth noting that both the {\em Fermi}-LAT
detections in 2008 November--2009 January and in 2012 July-October correspond
to periods of high optical activity of the NLSy1. This is another indication in favour
of an association between the $\gamma$-ray source and FBQS
J1644$+$2619. Unfortunately, as shown in Fig. \ref{Fig2}, no optical observations
simultaneous to the $\gamma$-ray peaks observed in 2012 August and October are
available. The average apparent isotropic (rest-frame) $\gamma$-ray luminosity of the source is
1.6$\times$10$^{44}$ erg s$^{-1}$, comparable to the value estimated for the
NLSy1 1H 0323$+$342 over the first 4 years of {\em Fermi} science operation
\citep{dammando13d}, but smaller than the other $\gamma$-ray NLSy1 by one to two
orders of magnitude.
The average photon index of FBQS J1644$+$2619 ($\Gamma_{\gamma}$ = 2.5 $\pm$
0.2) is similar to those observed in flat spectrum radio quasars as well as
$\gamma$-ray-loud NLSy1. No significant spectral evolution was observed during
the high $\gamma$-ray activity state in 2012 July-October, when the source
reached an average flux of (5.2 $\pm$ 1.0)$\times$10$^{-8}$ ph cm$^{-2}$
s$^{-1}$, and a daily peak of (66 $\pm$ 22)$\times$10$^{-8}$ ph cm$^{-2}$
s$^{-1}$, corresponding to an apparent isotropic (rest-frame) $\gamma$-ray luminosity of 1.9$\times$10$^{46}$ erg s$^{-1}$. 

In the radio band, the source has an inverted spectrum ($\alpha_r$ = 0.4 $\pm$
0.2, where $S \propto \nu^{\alpha_r}$) between 1.7 and 8.4 GHz \citep{doi11} and a very high core
dominance\footnote{The core dominance, $r$, is defined as the flux ratio of the core
  to extended emissions.} \citep[$r$ = 3.19;][]{doi12}, suggesting the
presence of a relativistic jet. Its brightness temperature of
7$\times$10$^{9}$ K, derived from the 1.7 GHz VLBI flux density of its unresolved core, is too high for free-free emission, indicating that non-thermal processes dominate the radio
emission. However, this value is lower than the limiting value
predicted by the inverse Compton catastrophe \citep[e.g.,][]{readhead94}, and in
contrast to what was observed in the $\gamma$-ray emitting NLSy1 SBS
0846$+$513 \citep{dammando13} and PKS 1502$+$036 \citep{dammando13c}, where
the variability brightness temperature exceeds 10$^{13}$ K.

On parsec scales the source has a one-sided core-jet structure \citep{doi11},
suggesting the presence of Doppler beaming effects. From
the core dominance, \citet{doi12} estimated a jet speed of $\beta$ $>$ 0.983 and
a viewing angle of $\theta$ $<$ 5$^{\circ}$, corresponding to a Lorentz
  factor $\Gamma$ $>$ 5. In the same way, a Lorentz factor $\Gamma$ $>$ 6.5
  and $\Gamma$ $>$ 2.2 has been estimated for the $\gamma$-ray emitting NLSy1 PMN J0948$+$0022 and 1H 0323$+$342, respectively. On kpc scale the source
shows a two-sided structure with extended lobes each dominated by a hot spot
near its outer edge \citep{doi12}, reminiscent of a Fanaroff-Riley type II radio galaxy.

FBQS J1644$+$2619 is included in the ROSAT All Sky Survey Bright Source
Catalogue with a 0.1--2.4 keV flux of 2.6$\times$10$^{-12}$ erg cm$^{-2}$
s$^{-1}$ and a photon index $\Gamma_{\rm\,X}$ = 2.03$^{+0.25}_{-0.28}$
\citep{yuan08}. On 2003 June 3 the source was observed by the {\em Chandra}
ACIS-S for 3 ks. The 0.3--5.0 keV spectrum was fitted with a power law with
$\Gamma_{\rm\,X}$ = 2.19 $\pm$ 0.17. A fit of comparable quality was
obtained adding a soft component (e.g., a black body) and the resulting photon
index was $\Gamma_{\rm\,X}$ = 1.8$^{+0.6}_{-0.3}$ \citep{yuan08}. As noted in
Section \ref{SwiftData}, a flat photon index ($\Gamma_{\rm\,X}$ $\sim$ 2) was
observed by {\em Swift}/XRT in 2011 December 26. In contrast, a relatively
hard spectrum ($\Gamma_{\rm\,X}$ $<$ 2) was observed in the other $\gamma$-ray-loud NLSy1, SBS
0846$+$513 \citep{dammando13}, PMN J0948$+$0022
\citep[e.g.,][]{dammando15,foschini11}, 1H 0323$+$342 \citep{dammando13b}, and
PKS 1502$+$036 \citep{dammando13c}. The relatively hard spectrum has been interpreted as an
indication for a significant contribution of a relativistic jet in X-rays. In
the case of PMN J0948$+$0022 the contribution of a soft
X-ray excess was evident below 2 keV \citep{dammando14}. 

The rather flat X-ray spectrum we have found in FBQS J1644$+$2619 may be due to the fact that the {\em Swift}
observation was performed when the source was in a low $\gamma$-ray state, and the
jet contribution to the X-ray spectrum could not hide the Seyfert component
(e.g. the accretion disc and hot
corona). Alternatively, the X-ray energy range observed by {\em Swift}/XRT may
cover the part of the spectrum including the tail of the synchrotron emission
and the initial rise of the inverse Compton emission, as in the intermediate-synchrotron-peaked BL Lac objects. A deep X-ray observation of
FBQS J1644$+$2619 with {\em XMM-Newton} will allow us to determine if the X-ray spectrum is
completely dominated by the jet emission or if there is some contribution from
the accretion flow (e.g., the soft excess, inverse Compton emission from
the hot corona above the disc or other reflection features such as the Fe K$\alpha$ line).

Since NLSy1 are thought to be hosted in spiral
galaxies \citep[e.g.,][]{deo06} the presence of a relativistic
jet in these sources seems to conflict with the paradigm that such
structures could be produced only in elliptical galaxies \citep[e.g.,][]{marscher09}. The most powerful jets are
found in luminous elliptical galaxies with very massive central black holes (BH) and low
accretion rates \citep[e.g.,][]{mclure04,sikora07}. This has been interpreted
as indirect evidence that a high spin is required for the jet formation, since
at least one major merger seems to be necessary to spin up the super-massive
black hole. At the same time, low accretion rates, which are usually associated with geometrically
thick advection dominated accretion flows, may be important in jet formation
by creating large-scale poloidal magnetic fields \citep{sikora13}. 
Using the relation between BH mass and broad line width, \citet{yuan08} estimated a BH mass of 8$\times$10$^{6}$ M$_{\odot}$
for this source, one of the smallest values among the 23 sources in the
sample. However, \citet{calderone13} pointed out that BH masses estimated for
NLSy1 by modelling optical/UV data with a Shakura and
Sunyaev disc spectrum could be significantly higher than those derived on the
basis of single epoch virial methods \citep[see
also][]{marconi08,decarli08}. In particular, for FBQS J1644$+$2619 they found
a value of 10$^{8}$ M$_{\odot}$. 
Even adopting this higher mass, FBQS J1644$+$2619 lies in the
low tail of the blazar mass distribution, challenging the theoretical scenarios
proposed for the production of relativistic jets. This object has the second
lowest redshift of the $\gamma$-ray NLSy1 (the most nearby object being 1H 0323$+$342), which makes it a good candidate for detailed studies of the host
galaxy \citep[e.g.,][]{leon14}. While the Sloan Digital Sky Survey image suggests a disc-like galaxy, observations with
higher resolution are needed in order to establish the morphology of the host
and thus to obtain important insights into jet formation and development. 

FBQS J1644$+$2619 was detected only sporadically in $\gamma$-rays during
2008 August--2014 December. This clearly demonstrates the importance of {\em
  Fermi}-LAT's continuous survey coverage for identifying variable
$\gamma$-ray emitting AGN like the
NLSy1. Detections of new NLSy1 like FBQS J1644$+$2619 by {\em Fermi}-LAT are
important to better characterize this new class of $\gamma$-ray-emitting AGN
and to understand the nature of these objects. Further multifrequency observations of this source from radio to $\gamma$-rays are needed to investigate in detail its characteristics over the entire electromagnetic spectrum. 

\section*{Acknowledgments}

The \textit{Fermi} LAT Collaboration acknowledges generous ongoing support from a number of agencies and institutes that have supported both the
development and the operation of the LAT as well as scientific data
analysis. These include the National Aeronautics and Space Administration and
the Department of Energy in the United States, the Commissariat \`a l'Energie
Atomique and the Centre National de la Recherche Scientifique / Institut
National de Physique Nucl\'eaire et de Physique des Particules in France, the
Agenzia Spaziale Italiana and the Istituto Nazionale di Fisica Nucleare in
Italy, the Ministry of Education, Culture, Sports, Science and Technology
(MEXT), High Energy Accelerator Research Organization (KEK) and Japan
Aerospace Exploration Agency (JAXA) in Japan, and the K.~A.~Wallenberg
Foundation, the Swedish Research Council and the Swedish National Space Board
in Sweden. Additional support for science analysis during the operations phase is gratefully acknowledged from the Istituto Nazionale di Astrofisica in Italy and the Centre National d'\'Etudes Spatiales in France.

Part of this work was done with the contribution of the Italian Ministry of
Foreign Affairs and Research for the collaboration project between Italy and
Japan. We thank the {\em Swift} team for making these observations possible, the
duty scientists, and science planners. The CSS survey is funded by the National Aeronautics and Space
Administration under Grant No. NNG05GF22G issued through the Science
Mission Directorate Near-Earth Objects Observations Program.  The CRTS
survey is supported by the U.S.~National Science Foundation under
grants AST-0909182 and AST-1313422. We thank J. L. Richards, L. Foschini,
Y. Tanaka, and the anonymous referee for useful comments and suggestions.

\end{document}